\documentclass[a4paper]{jpconf}
\usepackage{graphicx}
\usepackage{amsmath}%
\usepackage{amssymb}%
\usepackage{bm}
\usepackage{enumerate}

\newcommand{\bea}{\begin{eqnarray}}
\newcommand{\eea}{\end{eqnarray}}
\newcommand{\beq}{\begin{equation}}
\newcommand{\eeq}{\end{equation}}

\def \be{\begin{equation}}
\def \ee{\end{equation}}
\def \ba{\begin{array}}
\def \ea{\end{array}}
\def \bea{\begin{eqnarray}}
\def \eea{\end{eqnarray}}

\begin{document}
\title{ Phase diagram  of the spin-1/2 $J_1$-$J_2$-$J_3$ 
Heisenberg model on the square lattice }

\author{
 Philippe~Sindzingre$^1$,
 Nic~Shannon$^2$,
 and
 Tsutomu~Momoi$^3$
}

\address{
 $^1$
 Laboratoire de Physique Th\'eorique de la Mati\`ere Condens\'ee, UMR
 7600 of CNRS,
\protect\mbox{Universit\'e P. et M. Curie, case 121,
4 Place Jussieu, 75252 Paris Cedex, France}
\\  $^2$
 H. H. Wills Physics Laboratory, University of Bristol,
 Tyndall Ave, BS8-1TL, UK
\\  $^3$
 Condensed Matter Theory Laboratory, RIKEN, Wako,
 Saitama 351-0198, Japan
}

\ead{phsi@lptmc.jussieu.fr}

\begin{abstract}
We  presents the results of an extensive numerical study of the phase diagram of the 
spin-$1/2$, \protect\mbox{$J_1$-$J_2$-$J_3$} Heisenberg model on a square lattice, for 
both ferromagnetic and antiferromagnetic nearest-neighbor interactions $J_1$, 
using exact diagonalization with periodic and twisted boundary conditions.   
Comparison is made with published spin wave calculations.   
We show that quantum fluctuations play a very important role, changing 
both the extent and the wave vector of classical spiral phases, and 
leading to new quantum phases where the classical spiral states have 
a high degeneracy.     These include a new phase with small or vanishing spin-stiffness, 
in addition to known valence-bond-solid and bond-nematic phases.
\end{abstract}

\section{Introduction}
 The effects of quantum fluctuations on the ground-state is 
a major question in frustrated magnetism.
One of the most studied examples is the $J_1$-$J_2$ Heisenberg model
with spin $S=1/2$ on a square lattice with competing antiferromagnetic (AF) first and second 
neighbor exchange, where the nature of the non-magnetic ground-state
that occurs around the maximally frustrated point $J_2/J_1\sim0.5$
has been much debated.   Suggestions include valence bond crystals (VBC's) 
with columnar, plaquette, or mixed columnar-plaquette order, and spin liquid (SL) states.   
One approach to this problem is to include a third-neighbor exchange $J_3$, and 
a recent study of the $J_1$-$J_2$-$J_3$ model by Mambrini {\it et al.}
using exact diagonalization (ED) in full and reduced basis sets
found plaquette VBC order extending down to the $J_3=0$ limit
~\cite{Mambrini06}.

In past years, several classes of layered vanadate and cuprate compounds,
which are well approximated by a spin-$1/2$ Heisenberg model on a 
square lattice have come to the fore~\cite{mclmtmm00,kaul,kageyama}.
Some of these seem to be well described by a $J_1$-$J_2$ model
with AF $J_2$, and either FM, or AF~$J_1$.   However others have more
complex properties which hint at longer range interactions.   
This motivates us to consider $J_1$-$J_2$-$J_3$ models with both 
FM and AF~$J_1$~\cite{shannon06,sindzingre09}.


In this proceedings paper, we  report results of  further studies
of the $J_1$-$J_2$-$J_3$ model: 
\begin{equation} \mathcal{ H}=
  J_1 \sum_{\langle ij \rangle_1} \mathbf{S}_i\cdot\mathbf{S}_j
+ J_2 \sum_{\langle ij \rangle_2}
       \mathbf{S}_i\cdot\mathbf{S}_j
 + J_3 \sum_{\langle ij \rangle_3}
       \mathbf{S}_i\cdot\mathbf{S}_j
\label{Hamiltonian}
\end{equation}
that include now both the case of FM and AF $J_1$. 
We set $J_1=-1$ ($J_1= 1$) when $J_1$ is FM  (AF)
throughout.  
As before we  principaly used exact diagonalizations (ED)
on clusters of  $N=16,20,32,36$ spins with periodic boundary 
conditions (PBC), and $N=16,20,32$ spins with twisted boundary 
conditions (TBC), complemented by spin wave analysis following
previous calculations of Chubukov~\cite{chubukov_1984} and
Rastelli {\it et al.}~\cite{rastelli86,rastelli92}.

We find a wide variety of ground states, including both collinear
and spiral phases, and states with no conventional magnetic order.
For the case of FM $J_1$ our present phase diagram is a refinement
of that proposed in~\cite{sindzingre09}.   Further support has been found for the 
suggestion that, bordering the $S(q,q)$ phase 
for $0.25\lesssim J_2 \lesssim 0.7$, $S(q,0)$, spiral order is unstable against a 
new form of order.  For the case of AF $J_1$ we find a  
picture for the nature of disordered phase next to the $\mathbf{Q}=(\pi,\pi)$
slightly different  from the one proposed in~\cite{Mambrini06}~:
mixed columnar-plaquette VBC  order as proposed in~\cite{Ralko09}, 
and provide evidence for additional new phases at larger values of $J_3$.  
In this paper we limit ourselves  to the ground-state in zero magnetic field.  
The study of  magnetization plateaux  will be reported in a further publication.

\section{Ground-state phases}

\begin{figure}[tb]
  \vspace{-3.5cm}
  \includegraphics[height=11truecm]{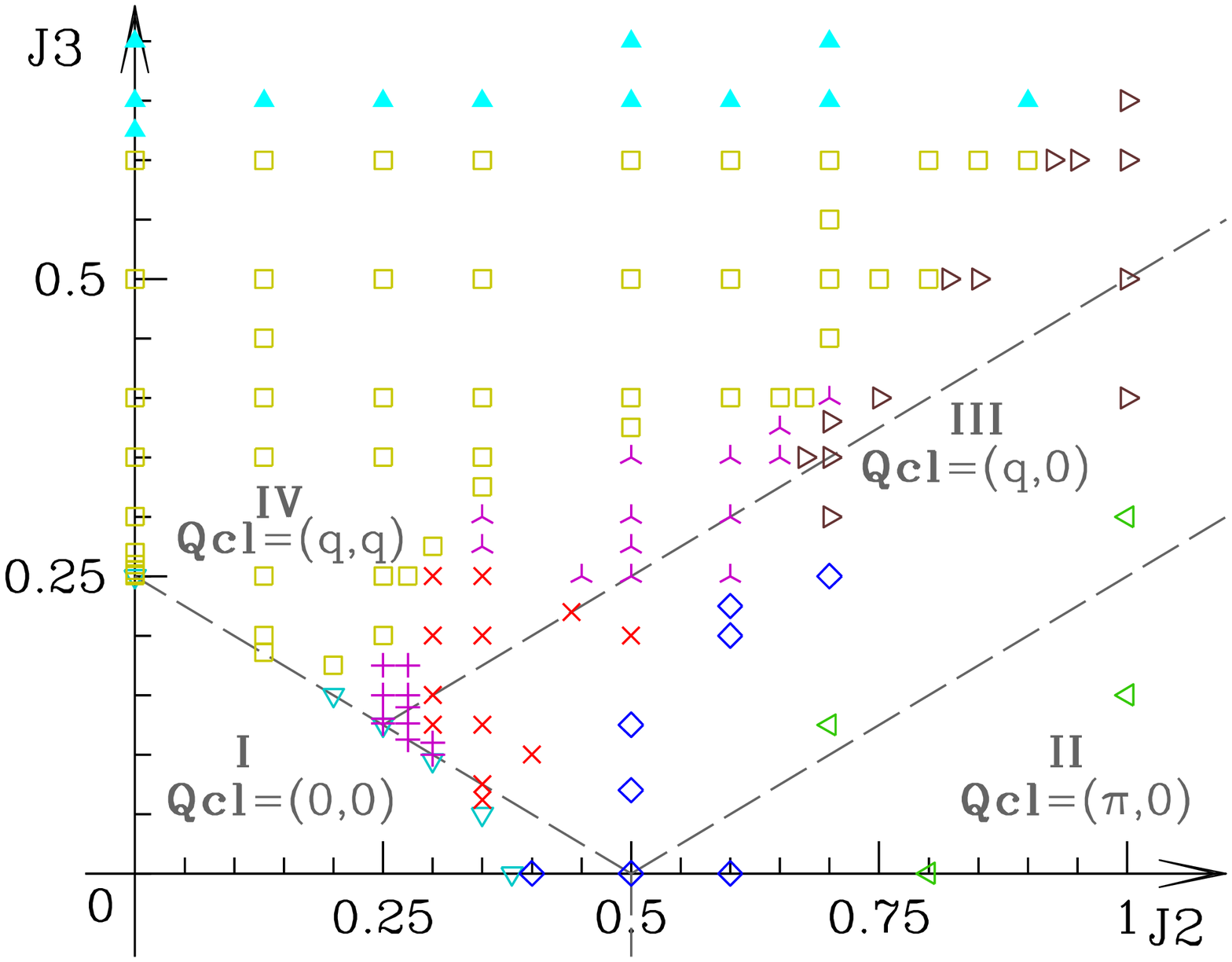}
  \includegraphics[height=10truecm]{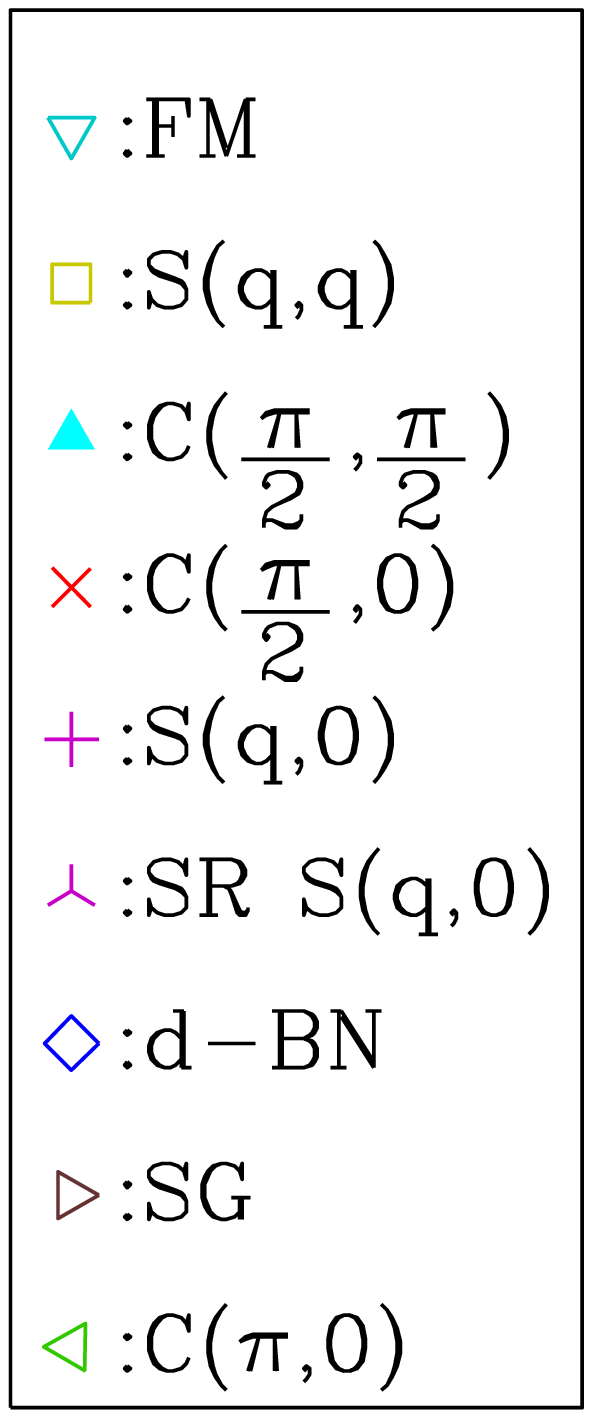}
 \caption{
(Color online).
Groundstate phase diagram for the spin-$1/2$, $J_1$-$J_2$-$J_3$ Heisenberg model
on a square lattice for $J_1 = -1$.  Symbols indicate the ground state found in exact diagonalization
of a 32-site cluster, allowing for twisted boundary conditions~:
FM~---~ferromagnetic; $S(q,q)$ or $S(q,0)$~---~spiral;
$SR\ S(q,0)$~---~sort-range order (see text);
$C(\pi/2,0)$, $C(\pi/2,\pi/2)$ or $C(\pi,0)$~---~collinear;
\protect\mbox{$d$-$BN$}~---~quadrupolar (bond-nematic) state;
$SG$~---~spin-gapped.
Dashed lines are the classical ($S = \infty$) phase boundaries.
\label{fig1}}
\end{figure}

\begin{figure}[tb]
  \vspace{-3.5cm}
  \includegraphics[height=11truecm]{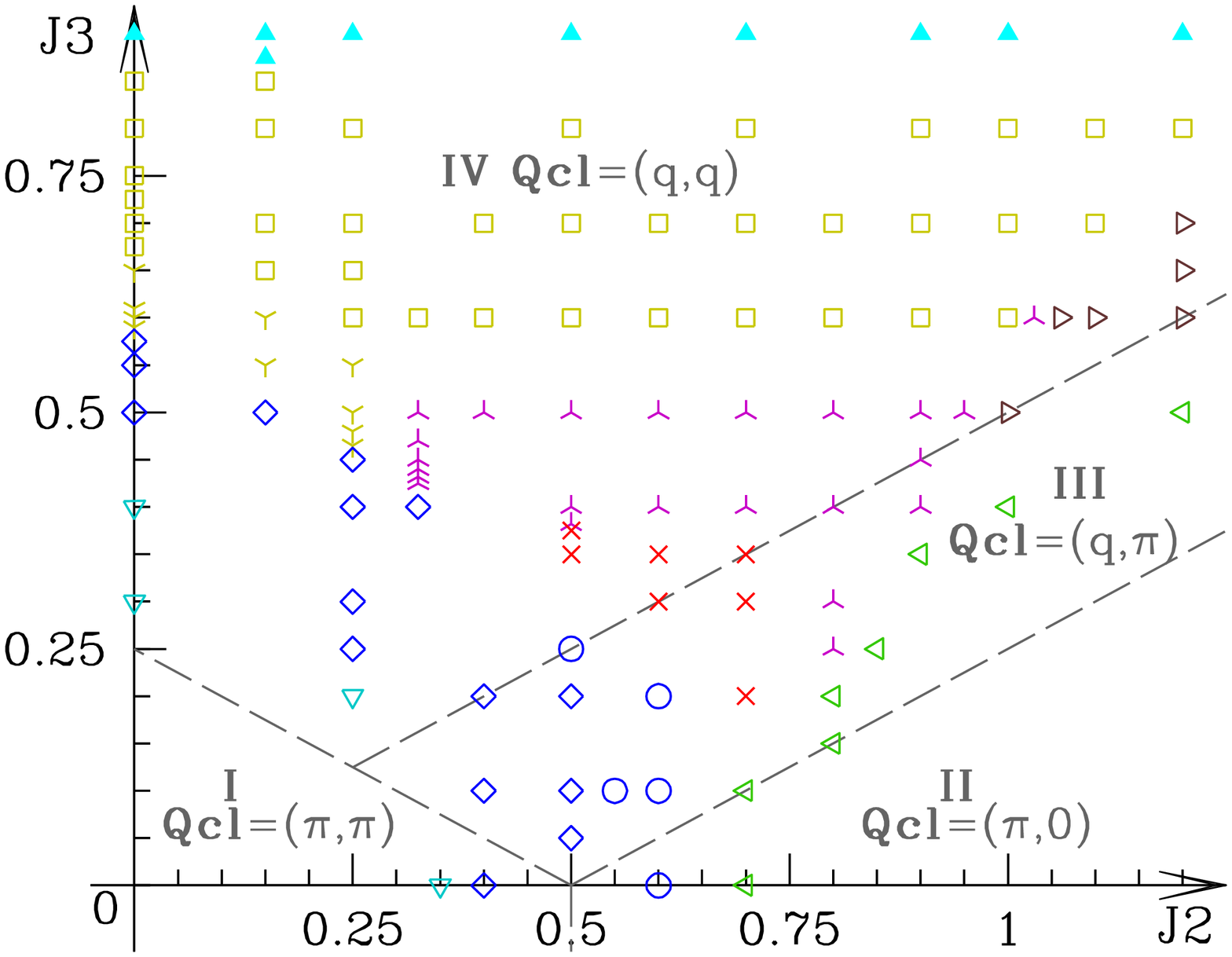}
  \includegraphics[height=10truecm]{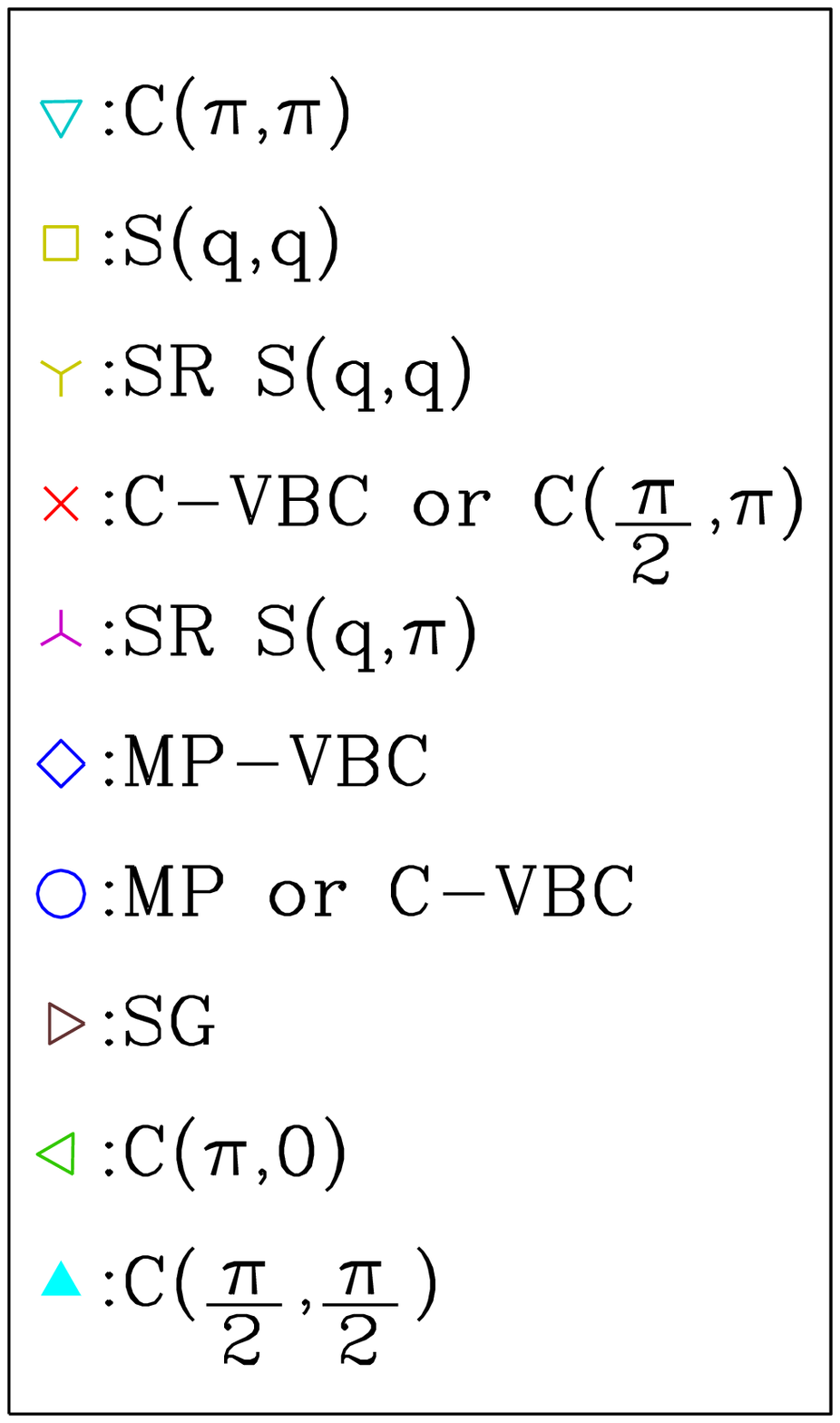}
 \caption{
(Color online).
Groundstate phase diagram 
for $J_1 = 1$.   
Symbols indicate the ground state found in exact diagonalization
of a 32-site cluster, allowing for twisted boundary conditions~:
$S(q,q)$~---~spiral;
$SR\ S(q,\pi)$~---~sort-range order (see text);
$C(\pi,\pi)$, 
$C(\pi/2,\pi)$, $C(\pi/2,\pi/2)$ or $C(\pi,0)$~---~collinear;
\protect\mbox{$MP$-$VBC$}~---~mixed columnar-plaquette VBC state;
\protect\mbox{$C$-$VBC$}~---~columnar VBC state;
$SG$~---~spin-gapped.
Dashed lines are the classical ($S = \infty$) phase boundaries.
\label{fig1_af}}
\end{figure}

Our main ED results for the spin-$1/2$ model are summarized in the phase
diagrams Fig.~\ref{fig1} (FM $J_1$) and Fig.~\ref{fig1_af} (AF $J_1$) 
in which the boundaries of the spiral phases have been deduced from
ED with TBC on $N=32$ clusters, and our conclusions on the nature of the
ground-state include the results of an analysis of the spectra obtained by ED 
of up to $N=36$ spins.   We note that these boundaries may differ from those
of an infinite system.

Classically, the $J_1$-$J_2$-$J_3$ Heisenberg model
on a square lattice has four magnetically 
ordered phases for $J_1$ FM or AF~\cite{rastelli86,rastelli92}.
These are collinear and spiral states at wave vector $\mathbf{Q}_{cl}$ which
extend over the regions labelled I-IV in Fig.~\ref{fig1} (FM  $J_1$)
and Fig.~\ref{fig1_af} (AF  $J_1$)~:
\begin{enumerate}[I --]
\item
a FM phase with $\mathbf{Q}_{cl}=(0,0)$ for FM  $J_1$,
a collinear phase with  $\mathbf{Q}_{cl}=(\pi,\pi)$ for AF $J_1$,
\item
a collinear phase with  $\mathbf{Q}_{cl}=(\pi,0)$
or $(0,\pi)$,
\item
a spiral phase with  $\mathbf{Q}_{cl}=(q,0)$ or $(0,q)$
where
$q=\cos^{-1}\left[-\frac{2J_2+J_1}{4J_3}\right]$ for FM  $J_1$, and
$\mathbf{Q}_{cl}=(q,\pi)$ or $(\pi,q)$ with
$q=\cos^{-1}\left[\frac{2J_2-J_1}{4J_3}\right]$ for AF  $J_1$,
\item
a spiral phase with  $\mathbf{Q}_{cl}=(q,q)$ or $(q,-q)$
where
$q=\cos^{-1}\left[-\frac{J_1}{2J_2+4J_3}\right]$
(so that  $q \in [0,\pi/2]$ for FM  $J_1$ 
and $q \in [\pi/2,\pi]$ for AF $J_1$).
\end{enumerate}
These magnetic orders will be referred as FM, 
$C(\pi,\pi)$, $C(\pi,0)$,
$S(q,0)$ $S(q,\pi)$, and $S(q,q)$ in this paper.

At a classical level, all transitions are continuous since
at the special point \mbox{$J_2 = 0.5$, $J_3 = 0$}, 
and on the boundary 
between III and IV, 
the classical ground state becomes degenerate with a family of spirals   
which interpolate continuously between the competing
values of $\mathbf{Q}_{cl}$, and appear as additional zero modes in 
linear spin wave theory (LSW) calculations. 
As a result of the high density of low-lying excitations,
LSW predicts that the sublattice magnetization of the classically ordered states
vanishes in the vicinity of 
{\it all} boundaries (except the boundary of the FM phase).  
Treated more accurately, quantum fluctuations may act to 
modify the wave vectors of spiral 
ground states~\cite{chubukov_1984,rastelli86,rastelli92},
enhance the extension of AF collinear phases, 
or lead to entirely new forms of order.  
The resulting phase transitions will generally be $1^{st}$ order.

\subsection{Phase diagram for FM $J_1$}
The phase diagram of Fig.~\ref{fig1} only differs from the one
presented in~\cite{sindzingre09} in the region 
\protect\mbox{$0.25 \lesssim J_2 \lesssim 0.7$} below the $S(q,q)$ phase
which we now label as $SR\ S(q,0)$ instead of $S(q,0)$.
There, as already noted in~\cite{sindzingre09},
the spectra of the $N=32$ cluster at the twist angle
that minimize the energy do not exhibit a well defined tower of state
caracteristic of spiral order.
Moreover, in this region, the energy 
is a very flat function of the twist angles over a large
region of twists which bridge between twists 
that allow to acomodate a $S(q,0)$ and a $S(q,q)$ spiral.
This region is indeed close to the line boundary between 
classical regions III and IV which is a line of 
degenerate spirals interpolating between $S(q,0)$ and $S(q,q)$ spirals.
Spiral order appears thus quite unstable and another kind
of ground-state occurs with short-range spin-spin correlations.
In view of the spectra, this ground-state does not exhibit 
a p-nematic order that correspond to the selection of a
preferential plane for the spins as found in~\cite{ldlst05}
and is probably spin-gapped.

\subsection{Phase diagram for AF $J_1$}

The phase diagram in the case of AF $J_1$ is shown in Fig.~\ref{fig1_af}.
The extension of the spirals is there also much reduced, 
and the $S(q,\pi)$ order is completely suppressed.
One has in addition a region, labelled   $SR\ S(q,q)$,
possibly spin-gapped, where the energy is minimized for TBC
but the spectra does not show spiral order.
This region is located between the VBC and $S(q,q)$ 
where it has been proposed that a  $Z_2$ spin-liquid occurs~\cite{cs04}.
The wave vector of the $S(q,q)$ spiral is also shifted by 
quantum fluctuations~\cite{sindzingre09}.
The $SG$ state is similar to the one found for FM $J_1$.
The collinear $C(\pi,0)$ and $C(\pi,\pi)$ phases clearly extends
beyond their classical boundaries.

The main difference is the appearence of a spin-gapped phase
with likely VBC order. 
In region adjacent to the $C(\pi,\pi)$ phase, this VBC
phase exhibit, in adddition to columnar dimer-dimer correlations,
strong plaquette correlations,
largest  around the line $(J_2+J_3)/J_1=1/2 $ and with decreasing $J_2$,
as previously pointed in~\cite{Mambrini06}.
However an analysis of the symmetry properties of the
low lying singlets present in the spectra  and their
evolution with the size of the clusters indicates
that an eight-fold degenerate mixed plaquette-columnar order,
as proposed in~\cite{cls00} and recently in~\cite{Ralko09}, or even
possibly a more complex pattern,
is much more likely than a purely plaquette VBC order. 
Moreover both the spectra and correlations
seem to favor a (quasi) pure columnar (four-fold degenerate) VBC order  
for $J_2 \gtrsim 0.5 J_1$.

 For $0.5 \lesssim  J_2/J_1 \lesssim 0.75$ and $J_3/J_1 \lesssim 0.35$,
we found a region where the spectrum that suggests 
(four-fold degenerate) $C(\pi/2,\pi)$ collinear order.
This state is one of the classical ground-states
at $ J_2/J_1 = 0.5 $ for $ 0 < J_3/J_1 <0.25 $,
degenerate with the $S(\pi/2,\pi)$ spiral.
But quantum fluctuations  may not be sufficiently strong
to stabilize this collinear state by an order by disorder
mechanism, at variance with the FM $J_1$ case,
where one nicely sees the signature of the similar $C(\pi/2,0)$ order.
In the AF case, such a collinear state seems in competion
with VBC order which could gain more from quantum fluctuations.
The ground-state may be spin-gapped with possibly columnar VBC
order, as suggested the correlations and the symmetry of the lowest 
singlets (which are the same as those of the singlets of the
Anderson tower of the $C(\pi/2,\pi)$ order) 
or even could be a spin-liquid.
The nature of the ground state in this region and the regions
labelled $SR\ S(q,0)$, $SR\ S(q,\pi)$ or $SR\ S(q,q)$ 
deserves further investigations.

\end{document}